# Rethinking the Generational Gap in Online News Use: An Infrastructural Perspective

Harsh Taneja (University of Missouri) [E]

Angela Xiao Wu (Chinese University of Hong Kong) [E]

Stephanie Edgerly (Northwestern University)

## Abstract

Our study investigates the role of infrastructures in shaping online news usage by contrasting use patterns of two social groups—millennials and boomers—that are specifically located in news infrastructures. Typically based on self-reported data, popular press and academics tend to highlight the generational gap in news usage and link it to divergence in values and preferences of the two age cohorts. In contrast, we conduct relational analyses of shared usage obtained from passively metered usage data across a vast range of online news outlets for millennials and boomers. We compare each cohort's usage networks comprising various types of news websites. Our analyses reveal a smaller-than-commonly-assumed generational gap in online news usage, with characteristics that manifest the multifarious

[E] Equal contributors (names in alphabetical order).

effects of the infrastructural aspect of the media environment, alongside those of preferences.

**Keywords**

Millennials, boomers, infrastructure, online news, news preferences, social media, legacy media, political polarization

**Rethinking the Generational Gap in Online News Use:**
**An Infrastructural Perspective**

This study explores the understudied role of infrastructures in shaping digital media use. Drawing heavily on self-reported survey data focusing on the preferences of various social groups, most existing literature assumes that, in a high choice media environment, content preferences are the principal determinant of media use. Our study differs from the norm both theoretically and methodologically. Theoretically, we explore the potential effects of the larger infrastructure of the online environment on people's web usage. We recognize that for users, infrastructural aspects remain largely invisible or taken-for-granted (Sandvig, 2013). Hence, to factor the infrastructural dimensions likely obscured in survey-based methods, our study takes a comparative approach to analyze passively collected metered usage data of social groups with distinct locations in the digital infrastructural environment. We compare two groups known for their particular content preferences that best enable us to discern the imprints of infrastructures on usage patterns.

Specifically, we use comScore data on shared usage between the 781 most popular news websites (along with two leading social network sites) for millennials (adults born after 1981), and boomers (born from 1946 to 1964) (Taylor, 2014). These two groups generally report vastly divergent news preferences and usage (Mitchell, Gottfried, & Matsa, 2015). Beyond comparing their aggregate usage of news websites, we employ a network approach to determine how each cohort actually uses various news websites in relation to one another. We compare online news use of the two cohorts on a variety of aspects that imply the role of content preference vis-a-vis that of news infrastructures. In particular, we focus on the infrastructural tendency for media consumption to display a power-law distribution, as well as the infrastructural legacy of media offerings. Our study reveals a generational gap in using online news much smaller than commonly assumed, and the generational differences and commonalities point to the understudied connections between digital media infrastructures and news usage. Our conceptualization about infrastructural effects, as well as our empirical approach to infrastructure through contrasting usage data, extends scholarship on media choice.

**Digital Infrastructures and Online News Usage**

Broadly speaking, the literature on media use has traditionally been divided into what we call the preference perspective and the infrastructural perspective. As a useful but rough guideline, the former focuses on meaning-making and the symbolic, with an emphasis on articulated pursuits of media, while the latter highlights to the format and materiality of the media system, regardless of content (Siles & Boczkowski, 2012).

The preference perspective, within which most existing mainstream understandings about news usage are situated, is rooted in the assumption that audiences are rational agents whose preferences—that is, enduring and self-conscious leaning toward specific types of media content—drive media choices (Owen & Wildman, 1992). These preferences are in turn due to their "tastes", "needs" or "genre preferences" (Webster, 2014). Studies following this paradigm find, for instance, that people's interests in news determine whether they seek or avoid news (Ksiazek, Malthouse & Webster, 2010), and their political ideologies determine the kind of news they gravitate towards, as well as systematically avoid (Stroud, 2011).

Stressing the insufficient explanatory power of individual preferences, an infrastructural perspective on media use enriches our understanding of online news usage. Given their bounded rationality, people rely on a variety of "media structures" to make their choices (Webster, 2014; Wonneberger, Schoenbach & Meurs, 2010). Invoking media structures in the study of media use is analogous to considering the media environment "infrastructurally", where a system's architecture is foundational to usage (Sandvig, 2013). For instance, TV channels and schedules constituted two important structures of traditional television infrastructure, and have historically explained a large part of people's viewing patterns (Webster, 2006). The infrastructural features of the digital media environment include delivery platforms, devices, algorithms, and access (Napoli, 2014).

Many recent studies have successfully integrated the two perspectives by factoring in both individual and structural determinants of media use (see Webster, 2014 for a review). They demonstrate that structural factors often moderate the role of preferences in explaining media use. For instance, many people otherwise uninterested in the news consume news in the presence of other interested viewers (Wonneberger, Schoenbach & Meurs, 2010). Likewise, availability moderates the gap between older and younger people's news use (Taneja & Viswanathan, 2014). Following this integrated approach, we identify two dimensions of digital news infrastructures that potentially operate alongside news preferences.

The first is the infrastructural tendency of the digital news landscape to exhibit a *power-law* use distribution, which we posit ultimately steers social groups towards relatively homogenous usage patterns, despite seemingly divergent preferences. Power-law distributions occur when a handful of outlets get the lion's share of attention, and the vast majority attracts relatively small audiences (Hindman, 2009). These skewed consumption patterns are common to any high-choice media environment regardless of the medium, platform or content genre. This is because how people's knowledge about others' media consumption often shape their own (a.k.a. positive network externalities). Increased access to such knowledge tends to further skew the power-law (Salganick et al., 2006). In the digital media environment, information discovery mechanisms such as search engines and 'trending topics' nudge people towards already popular products (Webster, 2014).[1] The concentration of attention in the few most popular products is especially pronounced on digital media, as information about media options and what is popular is highly visible and spreads rather easily (Webster, 2014).

One noteworthy infrastructural mechanism contributing to this tendency for power-laws is *social media curation*. We argue that, contrary to mainstream understanding, social media work to unify news use patterns of social groups that have similar substantial reliance on social network sites (SNS).[2] This is because for members of a local or even national community, the personal network structures through by which information comes have significant overlaps. On SNS, weak and diverse connections are more influential in information spread (Bakshy et al., 2012). People get information not just from their small, homologous network of strong ties, but more effectively from much larger diverse networks that often consist of friends of friends. ThusThus, individuals have massive overlapping information sources on SNS. In effect, a small number of opinion leaders and power users may drive information sharing for extensive SNS communities. On Twitter, a small number of trending topics influence the nature of information that circulates. These trends are geography based and all users see the same trending topics irrespective of their demographic such as age. Further, network structure of Twitter follows a power law like structure where a few star users are ubiquitously followed and control a large share of attention (Cha, Haddadi, Benevenuto, & Gummadi, 2010).

The second infrastructural dimension that affects use is the multifarious forms of *infrastructural legacy* associated with online news, which contributes to the usage difference between social groups that vary in their temporal engagements with the digital landscape. Infrastructures due to their enduring features often stabilize preferences to result in embodied, habitual patterns of media use (LaRose, 2010; Rosenstein & Grant, 1997). Thus, we need to approach new media use through a historical lens, considering it "constructed complexes of habits, beliefs, and procedures" embedded in established usage patterns (Marvin, 1990, p. 8). For instance, people with existing routines of consuming older, established media are likely to bring these routines online (Rosenstein & Grant, 1997). As such, visiting sites with legacy status is more about established habits brought online and not specific preferences for content.

**Using Online News: Millennials and Boomers**

To gauge how news use is connected to infrastructural features of the digital environment, we compare actual web use behavior of groups that are differentially positioned in terms of digital infrastructures. In particular, the dimension of infrastructural legacy is best discerned by examining generational differences in news use. Moreover, since the infrastructural tendency for power-laws works to homogenize usage patterns, its role would be most perceivable when comparing how groups use content types for which they are known to have the most disparate preferences. This helps ascertain whether content preferences are likely to result in divergent usage.

Following this rationale, we choose to contrast online news usage by millennials—adults born after 1981, and boomers—adults born from 1946 to 1964 (Taylor, 2014). By many accounts (drawing mainly from self-reports), these two age groups engage with digital news media rather differently. Born into the digital age, a growing number of millennials expressed little interest in news and political information and appeared detached from traditional sources of news (Mitchell et al., 2015). While majorities of every generation report getting news from the Internet (American Press Institute, 2014), younger generations appear different in the sources they turn to for online news. For example, 61% of millennials cite Facebook as their top source for political news, compared to only 39% of boomers (Mitchell et al., 2015).

Specifically, the literature suggests that news preferences regarding three content types most distinguish millennials from boomers. In terms of *geographic orientation*, younger adults are more mobile and cosmopolitan than their predecessors, and prefer loosely connected networks that are sustained through friendships and thin social ties, rather than geographically bounded local communities (Castells, 2001; Bennett, 2008). As for *topical focus*, it is argued that millennials have grown up without developing a preference for general news, instead preferring news that relates to their personal identities and topical interests (Bennett, 2008). Regarding *partisan leaning*, millennials are known to be more politically and socially liberal compared to past generations, as well as less defined by shared identities along formal political party lines (Taylor, 2014). Such differences may explain why conservative-voiced news use across television, radio, and online are especially popular among older adults, while liberal online news is more popular with a younger audience (Edgerly, 2015). If these news preferences indeed correspond with online news, we should observe divergence in how the two cohorts use news along the lines of geographic orientation, topical focus, and partisan leaning.

In contrast, based on our preceding discussion about news infrastructure, the two infrastructural dimensions of the digital news environment potentially affect both age cohorts' online news usage in different directions. First, we believe that due to the infrastructural tendency towards *power-law* distributions, both millennials and boomers end up consuming an overlapping set of ultra-popular news outlets. Specifically, our infrastructural consideration about social media curation suggests much lower generational difference. This is because leading SNS are popular among both millennials and boomer, which means both cohorts might get exposed to similar sources of news despite the differences between their immediate network connections.[3] Second, based on their legacy status, news sites may vary in their traction across generations, which contributes to existing generational difference. Specifically, millennials have been raised with certain media options proffered by the growth of cable television and the Internet, which boomers have lived a significant portion of their lives without (Zukin et al., 2006).

In sum, by closely examining the nature of the difference/commonality between the two generations along each of these aspects, we may effectively infer the role of infrastructures vis-a-vis preferences in explaining online news usage.

**Measurements of Online News Usage**

Most empirical studies of news usage invoke individual preferences as predictors, a choice often conditioned on employing surveys as a method to collect self-reported preferences. Self-reports of media use can be inaccurate for two reasons. First, people, and young adults in particular, tend to over-report news usage because it is socially desirable or because they are unable to recall the content they consumed (Prior, 2009). Second, surveys can only include a limited number of questions, oftentimes about a select number of popular outlets. A significant departure, our approach to measuring online news use (1) relies on passively measured behavioral data for a more accurate picture of media consumption, and (2) includes an expansive range of news outlets to construct a holistic account of the digital news infrastructures.

Rather than limiting our comparison to usage of outlets between groups in the aggregate, we examine online news usage through a relational or a network perspective (Webster & Ksiazek, 2012). To construct a usage network, we rely on shared traffic data between an expansive range of news websites. This enables us to gauge how people use websites in relation to other sites, and infer which websites are used more often in conjunction with other sites. The more often a site is used in conjunction with other sites, the more central it is within this network of media use. While prior studies have analyzed such networks in the aggregate for all audiences (Webster & Ksiazek, 2012), we construct a network of online news usage separately for millennials and boomers.

As already argued, digital news infrastructures may explain usage alongside content preferences. These connections need examination both at the level of the overall usage network and that of shared traffic between each website pair. To examine the infrastructural tendency for power-laws driven by a few universally popular websites, we conduct an overall networks' comparison. For the social media curation mechanism integral to this dimension, we introduce major SNS into each cohort's usage network and observe the resulting impact on website centralities. If the dimension of infrastructural legacy shapes usage, we expect a positive relationship between website pairs with similar legacy status and shared traffic in the usage network for boomers but not millennials. Hence our first research questions and hypothesis:

**RQ1:** How different is the online news usage network of millennials from that of boomers in terms of overall network correlation and website centralities?

**RQ1a:** What is the impact of including major SNS websites in each cohort's usage network on network centralities?

**H1:** In the usage network of boomers' but not in that of millennials, a website pair with similar legacy status has higher shared traffic than website pairs with dissimilar legacy status.

Likewise, if content preferences materialize into news usage, we expect websites with similar content focus to positively associate only for the group that prefers the particular content type. Such an association would be insignificant or negative for the group without those preferences. For instance, we may find that two politics-focused sites tend to have higher shared usage among boomers, whereas this association is insignificant in the millennials usage network. Conversely we may see a positive association between shared traffic and technology focused sites in the millennials network but such associations may be absent for boomers. Hence our broad hypothesis:

**H2:** Each content type for which a cohort is known to self-report a preference, website pairs with similar focus on this content type will have higher shared traffic.

Lastly, we select a set of major websites with varying partisan slant and closely compare how these two cohorts use these sites in conjunction. We thus pose this question:

**RQ2:** Do the usage networks of partisan sites for the two cohorts differ according to their known political leanings?

## Data and Method

**Sample**. We use passively collected metered data on Internet usage from comScore, which in the U.S., has about one million people under continuous measurement and records their Internet use across computers, tablets and cellphones. Each month comScore reports the aggregated usage of each website that their panelists accessed.

We downloaded unique visitor numbers for the entire list of online entities measured by comScore during April 2015, a typical non-summer month that was sufficiently removed from major political events. This included both news and other websites. That list yielded 22,714 properties (with unduplicated audiences) with a total of 31,542 entities (each entity could be a group of websites, a website or a subsection of a website). For instance, "Google Sites" is a level 1 entity (a property) under which YouTube.com and Google.com are level 2 entities. Google News and Google Maps are level 3 entities nested under Google.com.

First, we programmatically determined the level of each entity in the list. Second, as comScore labels each entity according to its genre, our script retained only those with the word "news" in their comScore category labels. This yielded 2,468 entities at various levels. Given that the audience of a higher-level entity include those of entities nested under it, many of the entities had duplicated audiences. Third, our script further removed this duplication by retaining the highest level for each such set of nested entities, yielding 1,739 entities. To avoid double counting, we manually checked this list to ensure that it contained only level per property. Fourth, we removed any entity that had less than 500,000 Unique Users to ensure we retain sites with reliable universe projections. Our final sample of 781 entities is very comprehensive; it includes legacy news sites (e.g., NY Times, CNN), news sections of major Internet portals (e.g., MSN News), and specialty digital-first outlets (e.g., Daily Kos, Vox).[4]

**Website Attributes**. Below, the first three attribute variables relate to news infrastructures whereas the last three to news preferences. To determine their values, we rely on existing data and classifications from two authoritative sources, comScore and Pew. This is combined with human coding, for which we recruited a graduate journalism student to determine the websites' legacy status and geography. To establish reliability, the authors and a trained coder first commonly coded a random sample of 10% of the websites.

***Website Popularity****.* We operationalized popularity of a website as its total monthly unique users for each cohort from comScore.

***Role of Social Media****.* We additionally include two prominent social media sites, Facebook and Twitter, to study their shared traffic with all the news sites in our sample.

***Legacy Status****.* We human coded a website "legacy" if its online operations were explicitly tied to physical news operations. Coders looked for certain features on the entity webpage that indicated a connection to another media platform (e.g., television/radio schedule, subscription to a print newspaper/magazine) and consulted Wikipedia to verify the history of the entity. Overall, 66% of entities in our sample were coded "legacy" (Krippendorff's alpha, hereafter "α"=.86).

***Geographic orientation.*** Geographic orientation was human coded in terms of local, national or foreign. Local sites included the websites of local and metro newspapers, local television stations, and some digital native local operations; these explicitly focused on a local area (DMA). National sites were US based organizations, such as the New York Times and Vox that focused on a national audience. Foreign websites were identified as those affiliated with organizations not headquartered with the US. Overall, 51% of entities were national ($\alpha$=.91), 36% local ($\alpha$ =.98), and 13% foreign sites ($\alpha$ = 0.86).

***Topical focus.*** For each entity comScore provides a broader primary category (e.g., news/information, entertainment, retail, services), and often a more specialized subcategory. We derived each website's topical focus from comScore's seven subcategories of "news/information," which included: business, entertainment, politics, technology, weather, general news, and newspaper. We combined the few sites with no subcategory with general news and newspapers, to form one category that we labeled "general" (511 sites). We retained comScore's codes for business (10.4%), entertainment (17%), politics (8%), technology (14.9%), and weather (2%).

***Subset of partisan websites.*** We followed past studies that have used audience composition as a proxy to determine political slant of news websites (for a longer discussion of this choice, see Flaxman et al., 2016, p. 6; Gentzkow & Shapiro, 2011). We used a recent Pew study (Mitchell et al., 2014) to select and categorize a subsample of news sites. This smaller sample contained only 24 news websites, with the determination of "neutral" (7 sites), "liberal" (13) or "conservative" (4) based on Pew's findings of audience composition. The full list of sites and their codes is provided in Table 3. We merged websites that occurred in our data as multiple entities (e.g., CNN Politics and CNN US) to make our subset fully compatible with the brands that Pew included.

**Measures and Analysis.** Our analysis of comScore's data focuses on a measure called "audience duplication," which captures the extent to which two media outlets (e.g., websites) are consumed by the same set of people in a given time period.[5] In a hypothetical universe of 100 people, if on a given day 20 people accessed both CNN and Facebook, the audience duplication between these two websites would be 20 or 20%.

We obtained pairwise audience duplication data for all the 781 websites, accessed during April 2015, separately for the comScore-provided age ranges of 18-34 years (millennials) and 55-64 years (boomers) in the United States. In total, the entire matrix for each age cohort contains the pair-wise audience duplication figures for 308,505 website pairs. We treat this matrix as a network with the websites as nodes and the extent of duplication between two websites as the value of the tie. Unlike prior studies that employ network analysis to study media use (Webster & Ksiazek, 2012) we refrain from dichotomizing tie values, which, preserves the granularity in our data.

From this data, we first analyze the news usage networks and compute *weighted degrees,* at the website level, which indicates its centrality by measuring the extent to which a website is used in conjunction with other sites. The higher the weighted degrees, the more central is the site. For cross group comparison, we normalized each generation's weighted degrees by dividing them with its total Internet user base. This normalized degree represents the total traffic per user a site shares with all other websites.

Next, we fitted a pair of regression models to discern the relative importance of different dimensions of news infrastructures and news preferences in determining each cohort's network— wherein ties are the dependent variable. In each model, we included three sets of independent variables, for which we transform website attributes into matrices based on pairwise similarities. For the attribute *legacy status*, for example, our independent variable is a matrix with cell value 1 if the row and column websites are both websites of legacy media organizations; else "0". Likewise, for the second independent variable, the *geographic focus* of each site, we computed three similarity matrices (respectively for local, national, and foreign), and for the third independent variable *topical focus*, six similarity matrices based on the six subcategories. Since network ties are not independent and identically distributed, we estimated p-values using quadratic assignment procedure, which calculates significance by simulating a large number of hypothetical matrices through simultaneous row and column permutations on the original matrix. Lastly, our analysis of usage of partisan-leaning sites focuses on the pairwise audience duplications between the subset of 24 partisan sites, a subnetwork we extracted from the full usage networks.

**Results**

**Network Descriptions**. In April 2015, comScore reported the total number of unique Internet users in the U.S. as 76 million for millennials (18-34) and 33 million for boomers (55-64). Overall, we find a high correlation between (r =0.86) between a site's visitor numbers for millennials and boomers. Similarly, the two network graphs are also highly correlated (r=0.76, p<.001 using quadratic assignment procedure at 1,000 repetitions), suggesting a large similarity in the usage networks of the two cohorts.

Table 1 contains the top 20 sites for weighted degrees for millennials and boomers. Overall, the two lists are quite similar with 14 of the top 20 sites common to both lists, though their relative ranks differ. Buzzfeed, for example, ranks 2nd among millennials, yet only 10th among boomers. Table 1 also points to some key differences. Several digital native sites that specifically target millennials such as Elite Daily, while Mic's mission is to provide a unique approach to news for a unique generation of millennials. Conversely, four out of the six sites exclusive to the boomer list have ties to television news (e.g., Fox News, CNBC) or newspapers (e.g., LA Times, Legacy.com).

[Table 1 About Here]

To further examine the "generational gap" in shared traffic, we ranked the differences in weighted degrees between cohorts for each news site. In Figure 1, values represent the additional amount of traffic per visitor that an individual site shares with all other sites, when comparing its millennial usage and boomer usage. Positive values thus indicate the site is more central in the millennial network than in the boomer network. We then zoom in onto the top 25 millennial sites and boomer sites, respectively, in terms of this generational gap. Corroborating the results highlighted in Table 1, we find that the websites that skew more toward millennials tend to be digital native sites that explicitly target millennials (e.g., Elite Daily, Mic), or specialize in entertainment (e.g., Hollywood Life, Perez Hilton), and technology news (e.g., Gizmodo, IFL Science). It is notable that Buzzfeed, while among the most central to both generations (Table 1), has the widest gap between its millennial and boomer usage. On the other side of the Figure 1 are the sites that skew toward boomers. Here legacy news sites predominate, which notably include six different Fox News sections, including Fox News entertainment. Ultimately, Table 1 and Figure 1 highlight several similarities among the millennial and boomer usage networks, as well as initial evidence of subtle differences between the networks thus answering RQ1. Our next set of analyses focusing on websites attributes investigate these differences more systematically.

[Figure 1 About Here]

**News Usage and Social Media**. Figure 2 reports the average weighted degrees before and after including each of the SNS for both cohorts. We find that while adding Facebook increases the average weighted degree of each age group in similar measures, the impact of adding Twitter is higher in the millennials' network. Further, for either cohort, adding these sites increases the average weighted degrees of both legacy media and non-legacy sites, however the increase is more pronounced for legacy sites (than non-legacy) in the boomer network. This answers RQ1a and suggests that users of legacy and non-legacy sites in either cohort are equally likely to use Twitter or Facebook.

[Figure 2 About Here]

**Network Predictions.** To test H1 and H2, we ran two regression models using similarities in news attributes as the independent variables to predict network ties in the millennial and the boomer network respectively. In both the regression models (results in Table 2), we logged the network ties as their distributions were positively skewed and taking natural logs made them relatively symmetric.

[Table 2 About Here]

First, we find that audience duplication between a pair of legacy websites tends to be higher (compared to that between a non-legacy pair or a legacy/non-legacy pair) for boomers, but not for millennial. Thus, H1 is supported. Second, compared to duplication between pairs of sites with dissimilar geographic focus, both cohorts have higher duplication between pairs of national sites and lower duplication between pairs of foreign sites. Notably, millennials tend to have a somewhat lower audience duplication between local sites, but the effect is insignificant for boomers. As for news topics, compared to website pairs with dissimilar topical focus, in the millennial network pairs of business and technology sites each have a higher duplication on average, whereas politics sites have lower duplication. For boomers, both business and weather sites have higher duplication, whereas the effects are insignificant for other news topics. Thus, H2 is partially supported.

**Partisan Subnetworks**. To answer RQ2, we selected a subset of 24 websites that prior research had examined in terms of audience political slant. We analyzed this subset separately for each age cohort. In Table 3, we report the weighted degrees of each website in descending order of their (normalized) values for each network. We find that, while neutral sites such as Yahoo News, CBS News and CNN are all among the highest for both cohorts, there are also interesting differences. For the millennial network, sites with high weighted degrees are more likely to be liberal. Buzzfeed, in particular, ranks the second highest, compared to its 9th position among boomers. In contrast, the top site for the boomer network is Fox News (which ranks only the 8th in the younger cohort). In fact, all four conservative sites rank higher in the boomer subnetwork.

[Table 3 About Here]

To further illustrate these differences, we visualized the two partisan subnetworks using ForceAtlas, a force-directed layout algorithm (Jacomy et al., 2014), which places nodes that have more weighted connections closer to one another. In Figures 3, nodes are news outlets, colored by partisan leaning, while ties are audience duplications between outlets, whose volumes are represented by thickness of the ties. The color of a tie combines the colors of its connected nodes. In the millennials network, liberal (blue) sites are in relative close proximity and largely sit at the center. The presence of thick green ties indicates sizable audience duplications between liberal and neutral (yellow) outlets. In the boomers' network, in contrast, liberal sites are more scattered while the conservative (red) sites, especially Fox News, take more central positions. In addition, we see thick orange ties connecting conservative and neutral outlets.

[Figures 3 about here]

**Generations Going Online: Infrastructural Convergence and Infrastructural Residuals**

To examine the role of infrastructure alongside preferences in shaping news usage, our study compares online news usage of two social groups—millennials and boomers. Overall, our analysis suggests that the millennials and boomers do not occupy exclusive enclaves of news websites. Rather, as the weighted degree distributions indicate, the usage-based networks for both cohorts are densely woven, indicating substantial audience duplications across our comprehensive sample of 781 websites. Our findings indicate that this infrastructural tendency of online news environments to exhibit power law like usage distributions, driven by universal usage of the most popular sites, leads to similarities in both generations' news usage patterns. As an instance, we found Yahoo-ABC News Network to be the most central site for both millennials and boomers, which means for users in either cohort, Yahoo-ABC News is the most used site in conjunction with whatever other sites they use, plausibly an artifact of Yahoo being among the most popular destinations online.

Relatedly, the sheer popularity of social media may have also contributed to more commonalities than differences in patterns of news usage between generations. Although counter-intuitive given the prevalent notion about the millennials' characteristic dependence on SNS (Mitchell et al., 2015), when analyzed as part of their news usage networks, Facebook and Twitter figure among the most central sites for *both* cohorts, a finding which supports our infrastructural expectation. Thus, both boomers and millennials use both these sites to similar degrees in conjunction with other news sites. This may be due to the distinct infrastructural capacity of these SNS to conveniently direct users to selected news content, a function that both generations have incorporated in their daily routines.

In short, we found that news infrastructures explain most of the usage patterns common to both generations. We call such infrastructural capacities to nudge various social groups toward similar engagements with media "infrastructural convergence."

Even the major generational differences detected in our analyses may be better explained by how each generation navigates the digital news infrastructure. To begin, compared to millennials, boomer news usage networks are more centered around legacy news organizations. At first glance this finding is consistent with the extant knowledge that boomers prefer legacy news because of an overall greater interest in hard news and politics. Although a closer look suggests that their greater reliance on legacy news online could be more of out of habit. Essentially, they bring their loyalties with legacy media to the online environment. For instance, we find Fox News far more dominant among boomers than millennials. The finding is unsurprising given the ideological bent of the news usage network. However, our data reveal that, not just "Fox News Politics" but all sections of Fox News are more central for boomers, with Fox News entertainment being the most central. Fox News' cable audiences are disproportionately loyal (Bennett & Manheim, 2006) and older than those of other cable news networks (Ariens, 2015). The legacy effect also shows in the much more central position of NDN (News Distribution Network) in boomers' network. NDN provides transmission architecture to help cable news outlets with their online distribution, and both parties (NDN and cable news) consequently acquire significant usage among boomers.

Among millennials, these legacy relationships are weaker; younger adults are more likely to have begun their relationships—not carrying over existing ones—with news organizations in the online environment. However, it is equally important to note that millennials are not shunning all legacy websites; many popular legacy websites (e.g., CBS News, CNN, Washington Post, New York Times) do occupy fairly central positions in their usage network. Several factors could be at play. First, many millennials still consume TV and radio, which is likely to bear on their online news usage. As of 2015, about half of millennial screen time in the US is spent on TV, and about half of the millennials self-report using TV as a news source (Nielsen & Sambrook, 2015). Second, if boomers are disproportionately loyal to legacy sites, many of which are popular "A-list" news providers, some millennials would derive positive network externalities from consuming them (e.g. on SNS).

Mainstream media sources may indeed gain more traction online than digital native media owing to their reputation, talent, as well as social and financial resources (Webster, 2014). But our empirical findings suggest that legacy media's online potential is heavily mediated by the users' generation and their relationships with these media on other platforms. Moreover, we found that the distinct preferences of boomers are not restricted to legacy media, but the "old timers" of the digital age as well. For example, the MSN brand (launched in 1995) takes a much more central role among the boomers, who probably developed reliance on MSN sites as their online life began with Microsoft's Explorer browser that comes with MSN as the default homepage. Today, millennials gravitate toward Buzzfeed (founded in 2006 as "The Media Company for the Social Age"), Mic (2011), and DailyDot (2011), while boomers still spend much time hanging out around MSN pages. In brief, compared to millennials, boomers' online news usage appears to be largely correlated with their existing habits formed prior to the prevalence of Web 2.0 or the Internet at large.

These differences cohere to suggest what we conceptualize as "infrastructural residual" in media consumption, a different type of infrastructural influence. The habituated usage stabilized under enduring media environments or infrastructures—or, news usage habits (LaRose, 2010)—may be imagined as an outcome of sociotechnical "imbrication," or the particular way in which human actors and the media-technological system have become entangled (Leonardi, 2012). Then when the media infrastructure is transformed, into one that is based online for instance, the new patterns of engagement, formed through a new process of imbrication, still bear some shape of the historical infrastructure. We call this the "residual" because, following Raymond Williams (1977), the residual elements of a past culture need to be filtered, "diluted," and "projected" as they are incorporated into the current dominant culture. The infrastructural residual in media consumption is most visible in places where the current media infrastructure resembles the earlier one. In our sample these manifest as legacy media websites as well as Web 1.0 portals, and unsurprisingly these explain most differences between cohorts. In this light, it is reasonable to imagine that in a future news landscape based on mobile access, millennials' news usage may exhibit a different pattern due to their own version of infrastructural residuals formed through desktop browsing experience and entrenchment of SNSs such as Facebook. In comparison, the younger, Generation Z's information intake may likely revolve around their favorite, more private social networking apps such as Snapchat and Whisper (Sparks & Honey, 2014).

Foregrounding the residual effect of historical media infrastructures is important to understanding online news usage because it highlights that the consumption patterns are not simply an immediate reflection of the users' viewpoints and lifestyle predispositions. Taking an infrastructural view here alerts us to move beyond symbolic content as the basis of media choice (which surveys are most apt to tackle), but also examine the latent yet enduring implications of the technical materiality of the media system, which tend to escape users' awareness. This analytical distinction is particularly crucial for examining news use during the technological changes of the digital age because we need to attend to the ways in which infrastructural residuals shape symbolic consumption (Sandvig, 2015). Consider how the Fox News Channel seems to carry its boomer audiences online to a habitat well crowded by various Fox sites, and how these online venues have become known for their conservative slant. The concept of infrastructural residual in fact allows the consideration of news preferences as partially endogenous to news consumption, in contrast to the common assumption about their exogeneity (Bowles, 1998).[6]

It could be argued that, without accounting for people's self-reported preferences, media usage data analyzed at the macro-level may overemphasize infrastructural effects. However, studies on media use analyzing respondent-level metered data have consistently found factors of larger media and social structures more powerful than individual traits (see Webster 2014). As demonstrated in our study, even without access to information solicited from individual respondents, careful research design with passively measured usage data can indicate the interplay between infrastructural features and content preferences.

**Conclusion**

We set out to discern the possible effects of the material, infrastructural aspect of the media environment, which usually occur in ways that elude users' articulation and even consciousness, by comparing news usage of two distinct social groups. Based on the specific locations of these social groups—millennials and boomers in this study—we discern how various infrastructural effects, alongside news preferences, play out in news usage.

Our study is a relational analysis of passively metered usage data across a vast range of news outlets. Compared to the commonly used survey method, this method reconstructs media usage in the high-choice digital era to a fuller extent. Contrary to prevalent accounts foregrounding divergent news preferences, our findings reveal only a modest "generational gap" in online news consumption. In essence, our analysis suggests that specific digital infrastructures complicate a purported direct relationship between news preferences and news usage.

In particular, "infrastructural convergence" explains why even in high-choice environments that offer users a great deal of autonomy to choose their preferred news sources, these two generations largely consume the same set of popular outlets. Infrastructural capacities, such as social media curation that affects choice-making and the rapid circulation of information on the Internet that amplifies social influences on media consumption, homogenize news usage patterns of different social groups. To account for the observed discrepancies, evidence at large points to what we call "infrastructural residuals," the specific manner in which users' past habituation under historical media infrastructures figures into their engagements with the present media environment. We also demonstrate cross-generational comparison as a fruitful site for further research on how news and broader infrastructures shape media usage. Although our study particularly highlights the effect of age in mediating usage within news infrastructures, the conceptual points we raise about the how news infrastructures shape usage can be adapted to other instances where preferences and infrastructures work in tandem.

York: Oxford University Press.

**Table 1**

*Top Sites by Weighted Degrees (Normalized and ranked for each cohort)*

| Millennials | | Boomers | |
|---|---|---|---|
| *Average Weighted Degrees (sd): 0.475 (0.836)* | | *Average Weighted Degrees (sd): 0.5 (0.805)* | |
| Yahoo-ABC News Network | 9.212 | Yahoo-ABC News Network | 8.069 |
| BUZZFEED.COM | 8.449 | CBS News | 6.013 |
| CBS News | 6.615 | CNN Brand | 5.477 |
| CNN Brand | 5.845 | NYTIMES.COM | 5.048 |
| WASHINGTONPOST.COM | 5.515 | NDN | 5.015 |
| NYTIMES.COM | 5.336 | WASHINGTONPOST.COM | 4.979 |
| People | 5.062 | TMZ | 4.452 |
| TMZ | 4.665 | People | 3.610 |
| FORBES.COM | 4.200 | MSN News | 3.609 |
| E ! Online | 3.963 | BUZZFEED.COM | 3.605 |
| WEATHER.COM | 3.873 | WEATHER.COM | 3.559 |
| ELITEDAILY.COM | 3.848 | NBCNEWS.COM | 3.530 |
| ACCUWEATHER.COM | 3.377 | ACCUWEATHER.COM | 3.482 |
| TIME.COM | 3.100 | FORBES.COM | 3.350 |
| Mail Online – News | 3.094 | Mail Online - News | 3.344 |
| NDN | 2.651 | Fox News Entertainment | 3.299 |
| NBCNEWS.COM | 2.593 | AOL-HuffPost Money & Finance | 3.260 |
| Entertainment Weekly | 2.568 | LA Times | 3.143 |
| IBTimes | 2.546 | CNBC | 3.089 |
| MIC.COM | 2.510 | LEGACY.COM | 2.966 |

*Note.* Highlighted entries are exclusive to each list.

**Table 2**

*Regression on Log(Audience Duplication)*

| | Millennials | | Boomers | |
|---|---|---|---|---|
| | *Estimate* | *Pr(>=\|b\|)\|* | *Estimate* | *Pr(>=\|b\|)* |
| (Intercept) | -2.423 | 0*** | -1.658 | 0*** |
| Legacy | 0.117 | 0.213 | 0.264 | 0*** |
| National | 0.520 | 0** | 0.471 | 0*** |
| Local | -0.262 | 0.012** | -0.087 | 0.380 |
| Foreign | -0.353 | 0.019** | -0.691 | 0*** |
| Entertainment | 0.211 | 0.243 | -0.056 | 0.750 |
| Business | 0.541 | 0.015** | 0.675 | 0.003** |
| General | -0.089 | 0.329 | -0.067 | 0.442 |
| Technology | 0.341 | 0.073* | -0.346 | 0.056* |
| Politics | -0.597 | 0.021* | 0.226 | 0.387 |
| Weather | 0.643 | 0.264 | 1.385 | 0.011* |

Note. Adjusted R –squares .04, .03; N = 780 nodes, 303810 undirected pairs (both models)

Significance codes: ‘*** 0, ‘**’ 0.05, ‘*’ 0.1.

**Table 3**

Weighted Degrees of Partisan Sites (Normalized and ranked for each cohort)

| Millennials | News Outlet | News Outlet | Boomers |
|---|---|---|---|
| 1.94 | Yahoo-ABC News Network | Fox News | 2.17 |
| 1.80 | BuzzFeed | Yahoo-ABC News Network | 1.86 |
| 1.44 | CBS News | Huffington Post | 1.62 |
| 1.41 | CNN | CNN | 1.50 |
| 1.23 | New York Times | CBS News | 1.40 |
| 1.17 | Washington Post | NBC News | 1.37 |
| 0.96 | NBC News | New York Times | 1.16 |
| 0.86 | Fox News | Washington Post | 1.11 |
| 0.62 | Huffington Post | BuzzFeed | 0.80 |
| 0.59 | BBC | BBC | 0.75 |
| 0.54 | USA TODAY | USA TODAY | 0.74 |
| 0.49 | Guardian | Guardian | 0.48 |
| 0.46 | NPR | TheBlaze | 0.46 |
| 0.44 | Slate | NPR | 0.42 |
| 0.40 | TheBlaze | Slate | 0.38 |
| 0.27 | New Yorker | Politico | 0.30 |
| 0.20 | Breitbart | Breitbart | 0.26 |
| 0.19 | MSNBC | MSNBC | 0.22 |
| 0.15 | Politico | Drudge Report | 0.19 |
| 0.11 | PBS | Bloomberg | 0.18 |
| 0.09 | Al Jazeera America | New Yorker | 0.14 |
| 0.04 | Drudge Report | Al Jazeera America | 0.14 |
| 0.03 | Bloomberg | PBS | 0.11 |

*Note.* Red: Conservative; Blue: Liberal; Yellow: Neutral.

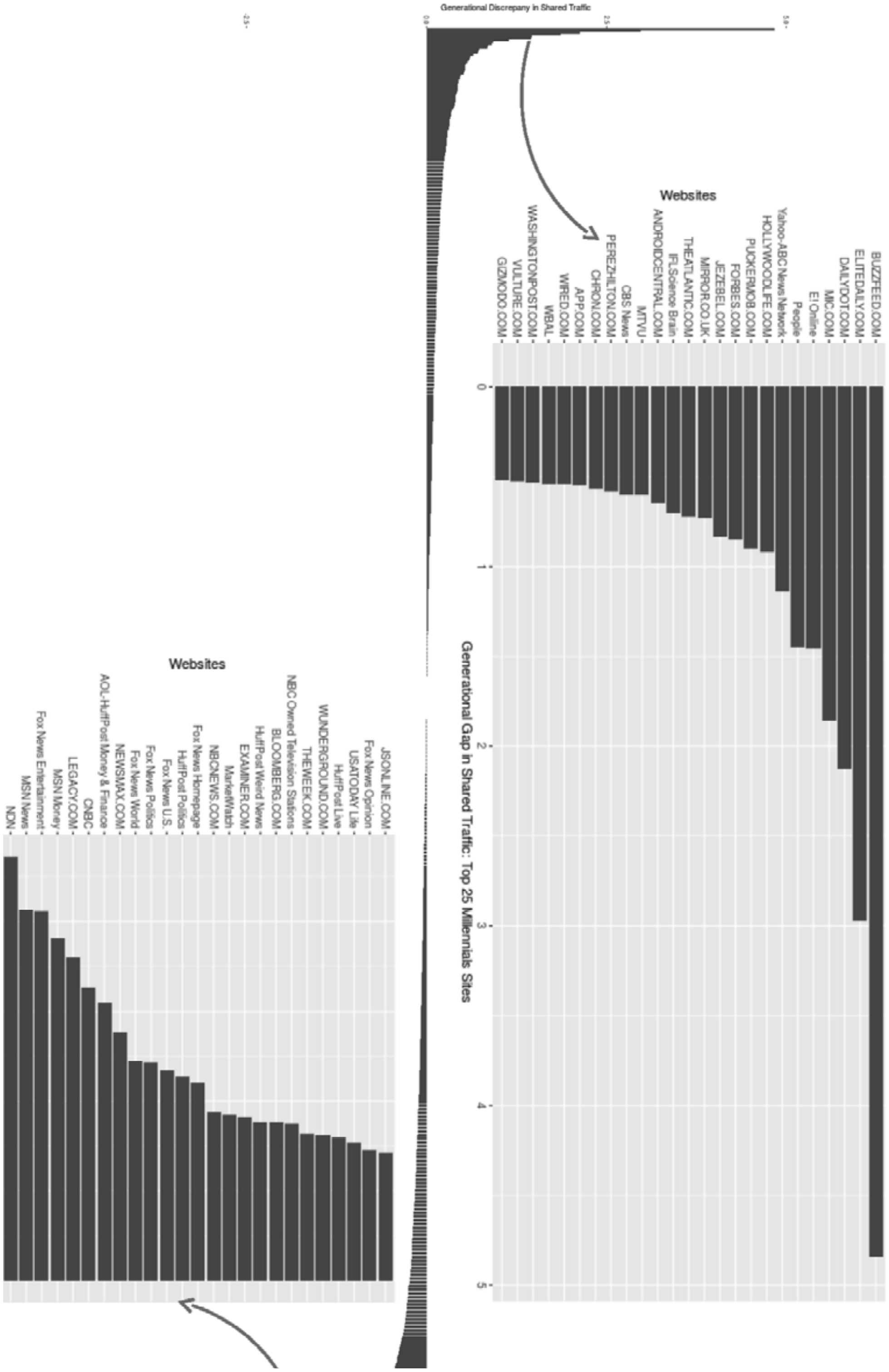

*Figure 1.* Generational Gap in Shared News Traffic.

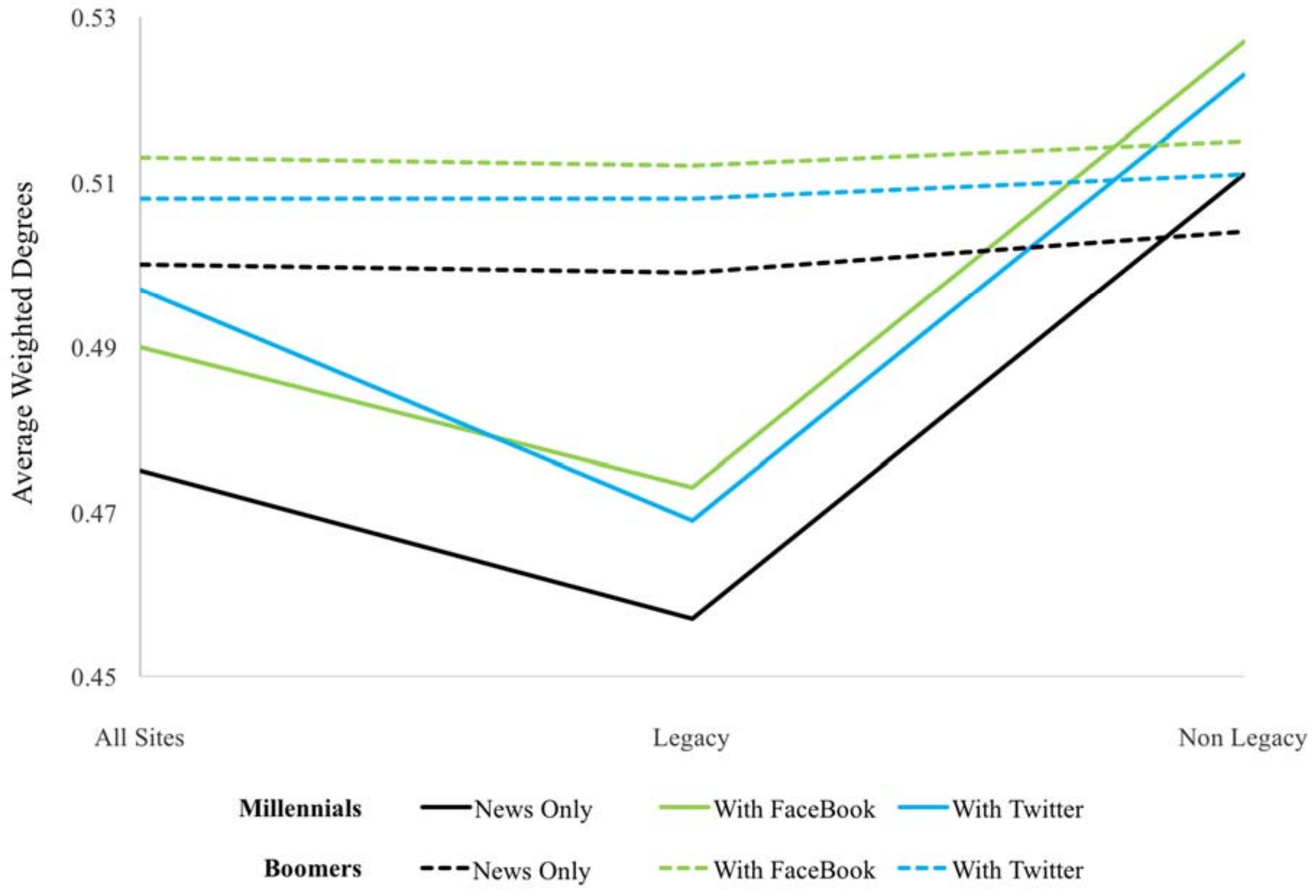

*Figure 2.* Effects of adding social media sites to online news usage networks.

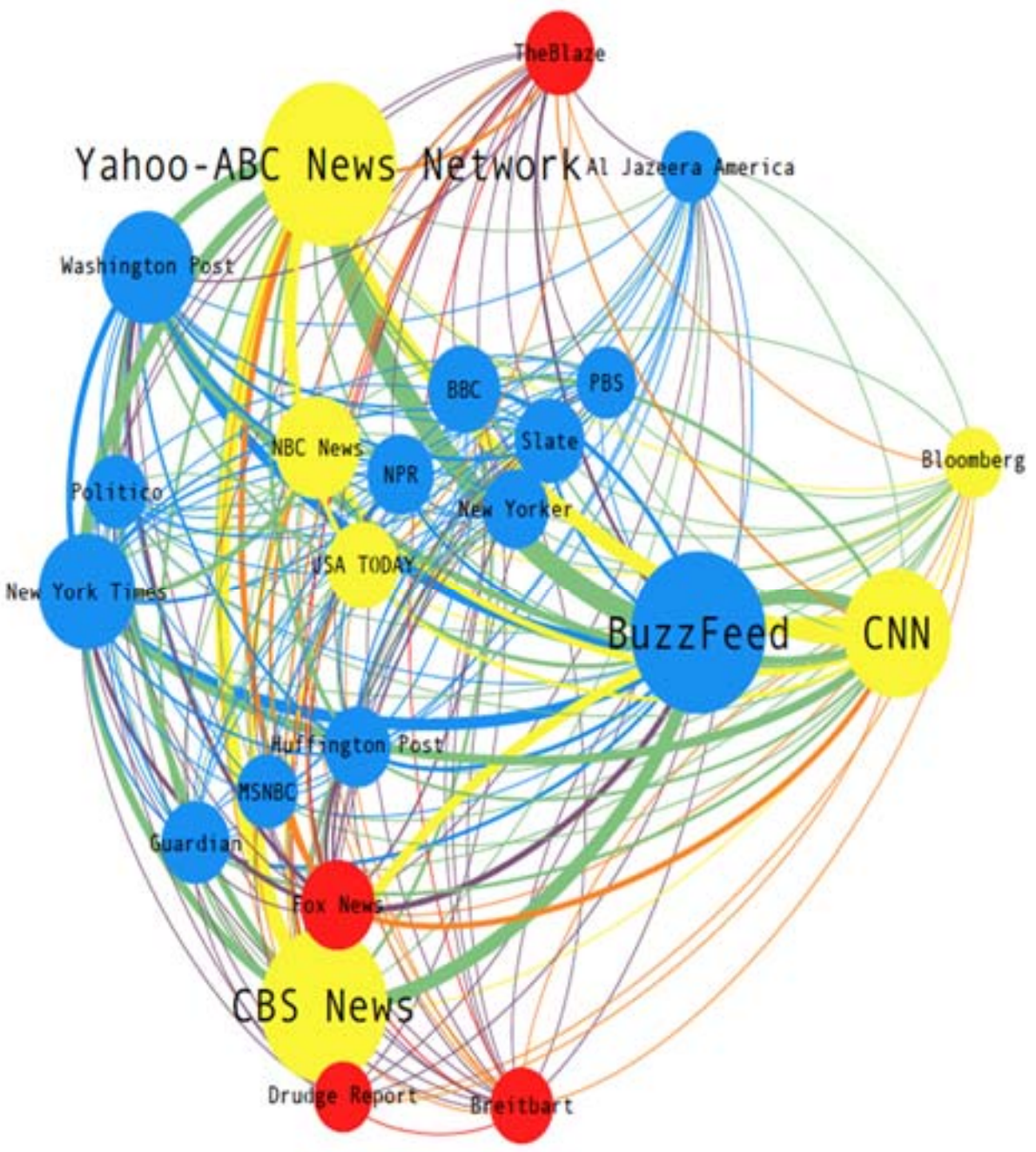

*Figure 3(a)*. Millennials' Subnetwork of Political Sites.

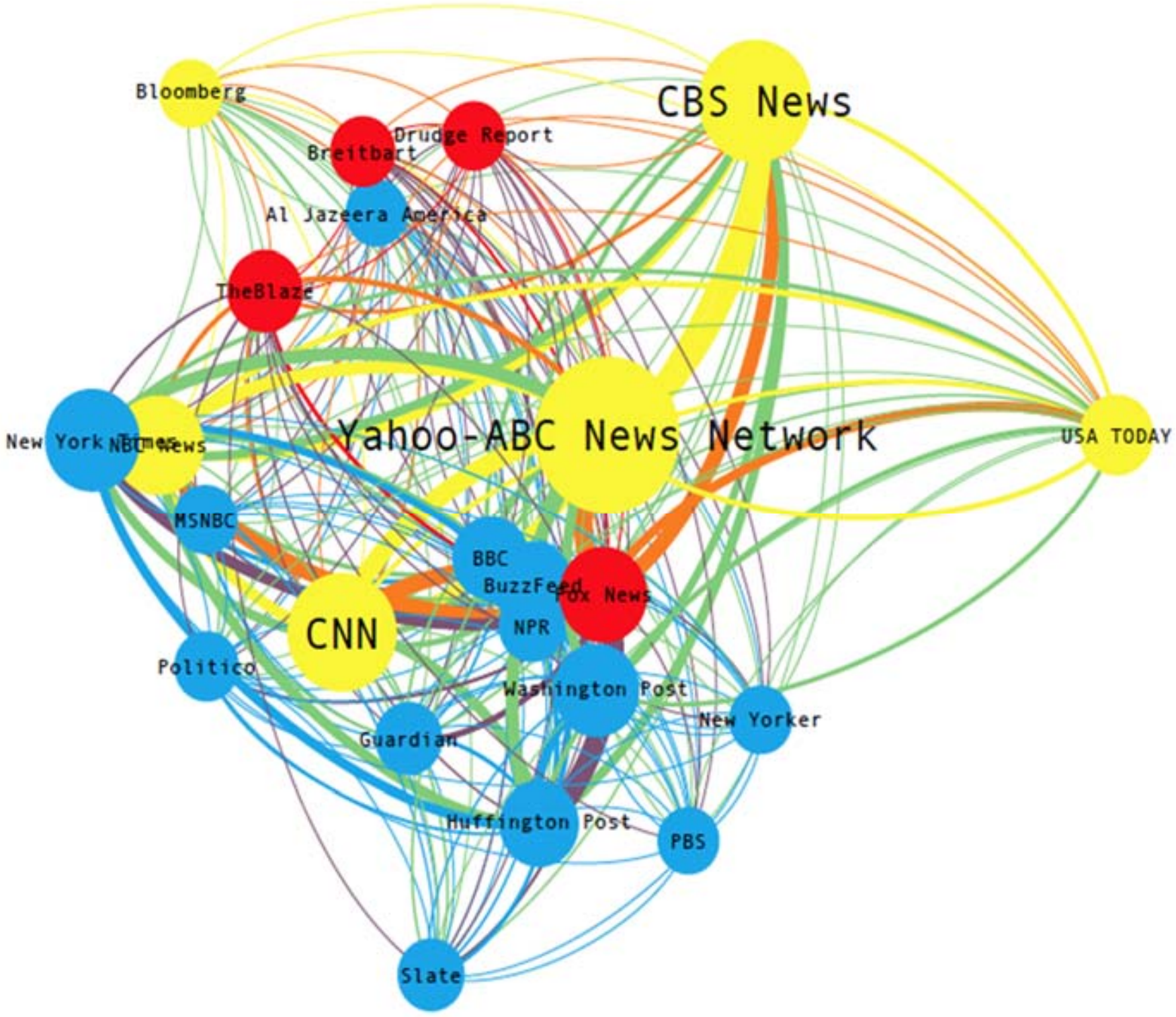

*Figure 3(b)*. Boomers' Subnetwork of Political Sites.

*Note.* Red: Conservative; Blue: Liberal; Yellow: Neutral.

Endnotes

[1] An alternative infrastructural explanation for imbalance in media markets is that "A-list" products have an advantage in attracting and retaining audiences because they generally have superior production values and higher marketing spends, regardless of content genres (Elberse, 2013).
[2] SNS increasingly serve as pathways for people to share, receive, and interact with news. In 2015, roughly two-thirds of users reported getting news from either Facebook (63%) or Twitter (63%) (Barthel et al., 2015). It was recently reported that Facebook surpassed Google as the top traffic source for news (Ingram, 2015).
[3] According to our comScore US data, 97% of millennials used Facebook and 54% used twitter in April, 2015, compared to 92% and 44% of boomers.
[4] Despite its comprehensiveness, our sample, by relying on comScore's classification, may leave out some non-traditional sources for news content, such as TheOnion.com, The Daily Show, and Youtube.com.
[5] This metric measures visits to websites but provides no detail on the duration and textual engagements.
[6] Most mainstream research on media choice in the social sciences, considers media preferences to be exogenous to use (e.g., Prior, 2007).